# Advanced Design of Self-Healing Dielectric Capacitors: New Universal Concept and Computational Method

Vitaly V. Chaban

Yerevan State University, Yerevan, 0025, Armenia. E-mail: vvchaban@gmail.com.

**Abstract**. A new computational method is herein discussed to systemize the development of new dielectric capacitor designs. The method predicts the identities and amounts of (1) gaseous products of decomposition, (2) the volume of the emerged solid phase, coined soot, (3) the band gaps of the soot samples, and (4) the electrical conductivity of the soot. The predictions can be made by using electronic-structure methods of atomistic simulations combined with the exploration of the potential energy landscape. The work discusses and rates the relative importance of each of the four introduced descriptors to unequivocally characterize possible capacitor designs. The concept is addressed to researchers working with enhanced energy storage designs and electrical engineers.

**Introduction**

Storage of energy in dielectric capacitors is based on a fairly straightforward concept. The equal opposite charges are induced on the electrode surfaces via an external electric current. The obtained configuration of matter is stabilized by a dielectric placed between the charged electrodes. As the system cannot relax to its initial state (possessing low potential energy), the capacitor remains charged, possessing high potential energy. The discharge begins after a useful load is connected to the capacitor's electrodes.[1-3]

The simple concept described above is nearly flawless, except for the possibility of an electrical breakdown. The breakdown represents an analog of lightning, which occurs in the dielectric when the latter is unable to withstand an appeared voltage. Non-uniformity of the matter arrangement, such as inappropriate contact between the capacitor elements, increases the chances of the breakdown to take place. The electrical breakdown generally ruins the capacitor.

Self-healing is possible in specific types of dielectric capacitors. Self-healing is a spontaneous phenomenon taking place after the breakdown and partially resolving the caused problems. The capacitor continues to perform but loses a part of its nominal capacity. Self-healing is presently known for the metalized film constructions of dielectric capacitors but may be potentially induced in other setups by tuning the chemical properties of the employed dielectric materials.[3]

In the present work, a new investigation scheme is explained step-by-step to develop advanced dielectric capacitors based on a profound understanding of the microscopic processes upon the electrical breakdown. First, the kinetic energy injection method is employed. This stage reveals the most thermodynamically stable components of the volatile by-products and solid phase remainder coined soot.[3-7] Second, the volume of the solid-phase sample is determined via the Monte Carlo integration over the surface of the obtained structures. Smaller soot samples designate higher-quality self-healing. Third, the band gap of the soot sample is determined. For that, the minimum-energy structural patterns are converted into infinite periodic systems. The inclusion of the Brillouin zones enhances the accuracy of the sampling. Fourth, the electrical conductivity is computed from the electronic structure of the soot samples.[8]

**Methods and Methodologies**

The kinetic energy injection method is widely employed nowadays to find various stationary points belonging to the potential energy landscape of a particular atomic assembly.[3-7] The method uses periodic kinetic perturbations of the point in the phase space to induce its movement to a different state. Kinetic perturbations following structural relaxations drive the system from its high-energy (unphysical) states toward its low-energy (physical) states through intermediate states. The kinetic energy injection method can be used to find realistic atomistic configurations starting from a nearly random or even fully random arrangement of atoms and molecules. This capability perfectly suits the problems related to the electrical breakdown, in which the temperature reaches subplasmic values. The PM7 Hamiltonian represents one of the most seamlessly suitable engines to compute interatomic interactions pertaining to the sampled potential energy landscapes.[9-11]

The volume of the soot sample can be computed by means of the Monte Carlo integration over the surface as soon as the optimized Cartesian positions of all soot atoms are determined. A general goal of the capacitor design is to obtain the minimum possible volume while maximizing the gas content after the electrical breakdown.

Because of the inherent limitations of the semiempirical methods, their predictions of the electronic structures of molecules and crystal structures are not considered ideal, unlike geometries and thermodynamic properties. The band gap of the soot sample can be better determined from plane Kohn-Sham DFT (PWKSDFT) calculations with van der Waals corrections. For that, the non-periodic soot structures are converted into periodic systems by implementing periodic boundary conditions. Both the geometries and lattice vectors are optimized to obtain the stationary point. The key assumption herewith is, that both PM7 and PWKSDFT provide accurate enough potential energy landscapes.[10-12] The low-energy structures obtained from PM7 must exhibit the same patterns as the low-energy structures obtained from PWKSDFT.

The electrical conductivity is obtained from the optimized periodic wave function, including 3×3×3 k-point mesh at a direct current. Specifically, the reported electrical conductivity is the average trace using the Kubo-Greenwood theorem.[8] The calculation takes into account intra-band, inter-band, and degenerate-band conductivity contributions by employing the Lorentzian representation of the Dirac delta function.[8]

**Results and Discussion**

This section exemplifies the analyses of the existing capacitor setups to assess their relative self-healing capabilities and outlines novel research directions to tune the properties of self-healing.

The capacitor consists of electrodes and a dielectric in between. Both of them respond to the electrical breakdown by losing their interatomic structures and evaporating. When the energy of the breakdown dissipates and the matter cools down, new chemical structures emerge. The elemental compositions of the new structures, including gases and soot, are limited by the initial elemental compositions of the electrode and dielectric. As identified by our research group, the elemental composition of the capacitor plays a paramount role in the quality of self-healing. In turn, the quality of self-healing determined the lifespan of a dielectric capacitor.

The electrodes of the commercial capacitors are produced from zinc-aluminum alloys, therefore, contributing Zn and Al atoms to the soot. These elements do not produce volatile compounds, so they contribute to the composition of the soot only. Zn is present in significantly larger molar fractions compared to Al, the latter being the core of the electrode. In turn, Al enhances the adhesion of the dielectric to the electrode thanks to the polarizability of the latter.

There is a relatively small number of dielectrics, which are employed in metalized-film capacitors. All of them are based on the carbon skeleton but somewhat differ in elemental compositions and structural descriptors. Polypropylene (PP), polyethylene terephthalate (PET), propylene carbonate (PC), polyimide (PI), polyphenylene sulfide (PPS), and polyethylene naphthalate (PEN) are presently considered in commercial technologies and academic research endeavors. PP, PC, and PET are deemed to be the most successful dielectric materials to implement the insulation in the metalized-film dielectric capacitors. PI, PPS, and PET exhibit encouraging resistance parameters and acceptable dielectric losses, yet their lifespans are inferior to those of PP-based and PET-based devices. We unraveled that the fraction of the oxygen atoms in the capacitor was essential for attaining higher-quality self-healing. In turn, the positive role of the hydrogen atoms requires more in-depth reconnaissance. This is because hydrogen tends to form dihydrogen gas at high temperatures instead of binding the carbon atoms of the soot to decrease the size of the latter. Other non-metal chemical elements can have a drastic impact on self-healing. Finding the most optimal elemental compositions can open novel avenues to engineer high-performance self-healing dielectric capacitors.

The electrical breakdown liberates a substantial amount of energy. At the epicenter of the arc discharge, the temperature reaches roughly 7,000 K, a subplasmic temperature. Herewith, the matter of the electrode and the dielectric vaporizes and all covalent bonds dissociate, irrespective of the actual activation energy barriers. As the energy dissipates, the atoms self-assemble into alternative chemical structures. Unfortunately, the previously vaporized atoms cannot self-assemble back to the dielectric polymers and metallic electrodes because those do not represent the global minimum state of the considered atomic ensemble. In the meantime, the atoms self-assemble into their low-energy structures following the real-time evolution of the respective thermodynamic potentials. Due to the very high kinetic energies of these species, the activation barriers of specific chemical reactions are not limiting factors. According to the described understanding of the process, it is possible to forecast the chemical identities of the self-healing products by finding the lowest-energy atomic configurations of the initial chemical compositions.

The first step of the introduced method is to find the lowest-energy atomic configurations corresponding to the initial elemental compositions of the polymers and electrodes. The kinetic energy injection method, described and validated elsewhere,[1,13-14] has been successfully used to achieve this goal. Figure 1 depicts the standard formation enthalpies for various chemical compositions of a capacitor.

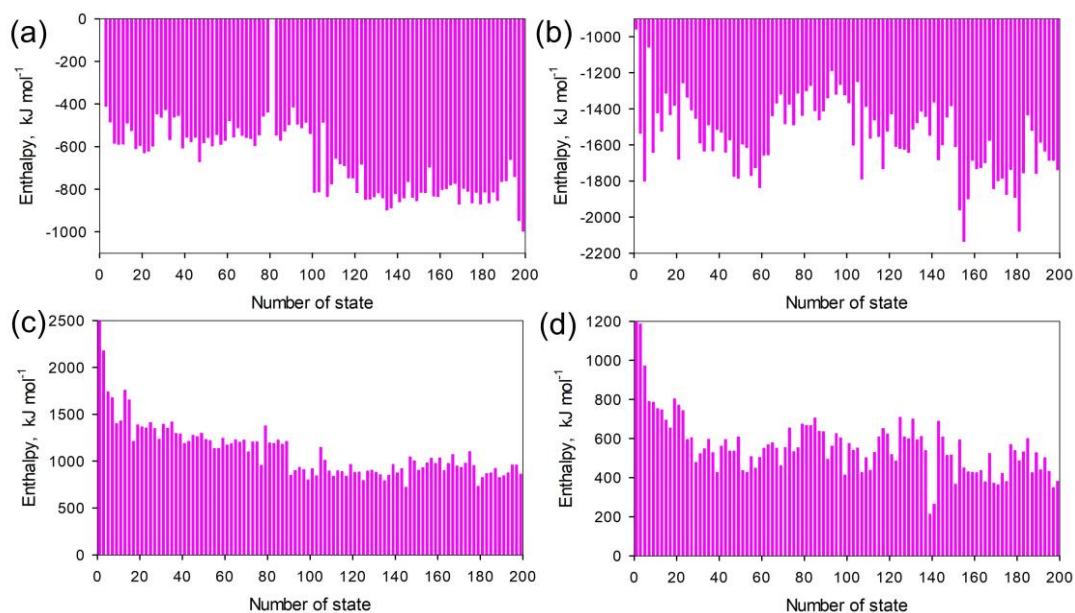

Figure 1. The formation enthalpies of various systems composed of model gold-electrode dielectric capacitors, in which the polymeric dielectric materials: (a) PP, (b) PET, (c) PC, and (d) PI.

The analysis started with random atomic arrangements. This is why the first located stationary points possess high enthalpies. Such stationary points, despite being valid, do not characterize the kinetically stable geometries of the system and may be safely disregarded. After several iterations of the kinetic energy injection procedures, the low-energy stationary points are detected. After that, the systems retain their general atomistic arrangements and jump among minimum states, one of which is the global minimum one. It is easy to notice that quite a few low-energy states exist for every chemical composition. They frequently differ from one another by some dihedral angles, yet rarely differ by valence angles, bond length, and atomic patterns. Mind that the obtained sets of the local minima contain duplicate geometries. Low-energy states, particularly, tend to be detected two or more times. The frequencies of stationary state detection, along with their potential energies, reflect their importance for the macroscopic system. The sign of the formation of enthalpies, as depicted, is inherent to each system and should not be used for their mutual comparison. Within PM7, the enthalpy of formation stands for the formation of the obtained atomistic configuration out of simple substances, i.e., the conventional thermodynamical definition is applied. Since the simulated polymers possess different chemical compositions, the reference states for the calculation of the formation enthalpies are accordingly different.

The predicted identities of the products, e.g., gases, may be validated through far- and mid-infrared spectroscopy. The covalent bonds maintaining the solid phase also provide their fingerprints in the FTIR spectra of the soot. Consequently, the computational results may be interfaced with in-lab physicochemical analysis, enhancing the credibility of the derived conclusions. Employing in-silico research to guide materials design is a particularly powerful research avenue.

The volume of the soot represents a paramount descriptor of the quality of self-healing. Smaller soot samples are less likely to be arranged as an inter-electrode semiconductive bridge. The engineering of the dielectric capacitor must strive to minimize the potential size of the soot sample emerging after the electrical breakdown. This can be achieved by enhancing the fraction of the gaseous by-products. The latter depends exclusively on the elemental composition of the capacitor as a whole. Various additives can be useful to tune the elemental composition to the desired values, whereas simulations can suggest interesting solutions to boost the formation of the low-molecular by-products. The volume of the soot can be obtained computationally by integrating over the surface of the solid phase.

The Monte Carlo method represents a powerful technique for approximating the volume of complex shapes. It works by randomly sampling points within a known volume that encloses the object of interest. By counting the proportion of points that fall within the object, one estimates its volume. First, a simple shape, like a cube or sphere, that completely encloses the object, is identified. Its volume is computed analytically. Second, a large number of random points within the enclosing volume is generated. Points should be uniformly distributed. Determine for each random point, whether it lies inside or outside the object. This can be done through ray casting or analytical geometry. Eventually, one calculates the ratio of the number of points inside the object to the total number of points and multiplies this ratio by the volume of the enclosing shape. The accuracy of the sought volume improves with a larger number of random points. For highly complex objects, techniques like importance sampling improve the efficiency of the evaluation.

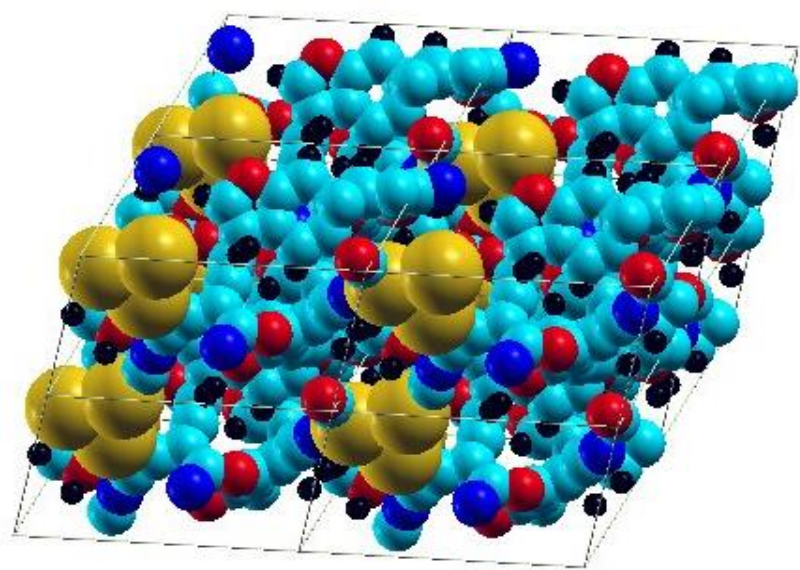

Figure 2. The periodic model of the soot sample was used to derive the band gap and electrical conductivity. Two unit cells are depicted herein to highlight the periodicity of the simulated solid.

As soon as the low-energy soot sample is predicted and its volume is computed, the electrical properties must be characterized. An ideal soot sample would exhibit the behavior of an insulator like its precursor polymer (PP, PET, PC, PEN, PPS, PPI). Recall that a major fraction of the soot is composed of the atoms previously belonging to the insulator. The major chemical element of the currently used polymers is carbon. Hydrogen atoms saturate the carbon's angling bonds but their role in terms of mass and volume is marginal. A relatively insignificant molar fraction, roughly two percent, comes from the metal electrodes. In turn, a major molar fraction of the electrode is composed of the zinc atoms in common capacitor setups. The bottom line is, that a basic model of the soot sample must contain carbon and hydrogen atoms, sporadically substituted by zinc atoms. Hydrogen atoms saturate carbon atoms only partially. Unsaturated carbonaceous structures are known to exhibit semiconductive and even conductive properties. The goal of the research is to detect the components of the soot, which are able to quench the electrical conductivity of the sample.

The other chemical elements, such as nitrogen, oxygen, and sulfur, tune the electronic structures of the soot samples somewhat. Figure 3 exemplifies the band gap of the soot obtained from the PET dielectric and gold electrodes, whereas Figure 4 reports the band gap of the PC dielectric with the same electrodes.

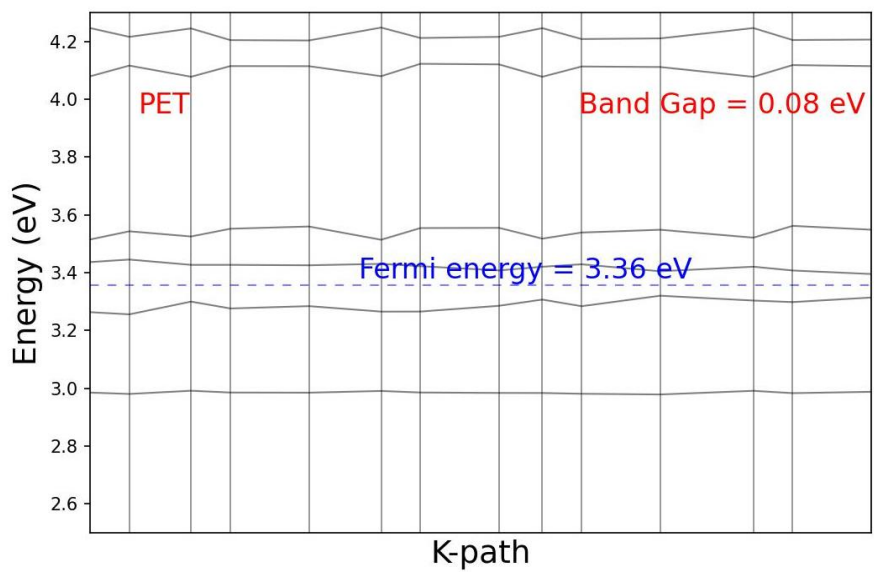

Figure 3. The valence band, the conduction band, and Fermi energy along the K-path of the simulated soot sample obtained from the PET dielectric.

Despite the differences in band gap and Fermi energy, the electrical conductivities of both samples are similar amounting to ~$2\times10^3$ S/m. This is because electrical conductivity depends on a few descriptors of the electronic structure, whereas the band gaps only reflect the energy difference between the valence and conduction bands. All electrical conductivities computed for various insulators (PP, PET, PC, PEN, PPS, PPI) and various electrodes (Zn, Al, Au) – not yet published – turn out to be between $10^3$ and $10^4$ S/m. These results position them as strong semiconductors according to the well-accepted scale. Compare this to pure metals (Al, Zn, Au, Cu) exhibiting much higher conductivities than $10^4$ S/m. The formed soot is less conductive than true conductors and significantly more conductive than true insulators.

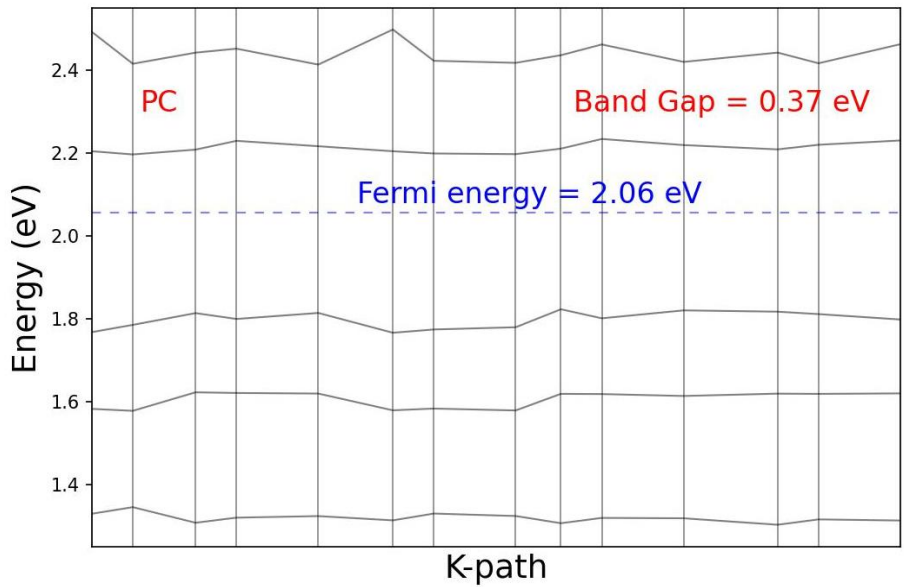

Figure 4. The valence band, the conduction band, and Fermi energy along the K-path of the simulated soot sample obtained from the PC dielectric.

Generally, a smaller band gap leads to higher conductivity, as electrons can more easily be excited from the valence band to the conduction band, where they can freely move. Yet, other factors are also very essential. A higher density of states at the Fermi level increases conductivity, as there are more available states for electrons to occupy and move. The effective mass of charge carriers, electrons and holes, determines their mobility. A lower effective mass allows charge carriers to move more easily under an applied electric field, leading to higher conductivity. The mean free path designates the average distance a charge carrier travels before colliding with an impurity, defect, or phonon. The thoughtful impregnation of impurities may allow to deterioration of the electrical conductivities of the soot samples. A shorter mean free path reduces scattering and decreases conductivity. The number of electrons and holes per unit volume also matters, whereas a higher charge carrier concentration fosters electrical conductivity. Impregnating large atoms into the admixture soot is an option to suppress the soot conductivity in addition to breaking its symmetry.

**Conclusions**

The transparent method has hereby been introduced to trial various compositions of the dielectric capacitor electrode. The method provides reliable computational predictions for known capacitor setups and can be validated through spectroscopic techniques, such as Fourier-transform infrared spectroscopy. The method can be applied to hundreds of potential capacitor setups, including dielectric materials, electrode materials, and additives, to evaluate their prospects in the context of self-healing. All stages of the proposed methodology are computationally affordable. The introduced vision greatly extends the scope of possibilities to develop better and cheaper energy-storage setups.

The materials science for dielectric capacitors is a largely undeveloped field, which does not go beyond combining a few materials for the electrode and half a dozen materials for the dielectric. This paper provides the first-in-history concept to automate the research schedule and attain the limits of capacitor lifespan and stability via optimizing self-healing.

**Conflict of Interest**

The author hereby declares no financial interests and professional connections that might bias the interpretations of the obtained results.

**Acknowledgments**

The author is presently an invited foreign professor at Yerevan State University. The author thanks Nadezhda Andreeva at Peter the Great Saint Petersburg Polytechnical University for preparing the used artwork that documents the method's workflow and performance.